\documentstyle[prl,epsfig,aps]{revtex}

\def\vS{{\bf S}}

\def \bi{\bibitem}

 \def\(({\left(}
 \def\)){\right)}
\def\bi{\bibitem}

\def \b{\beta}

\def \beqna{\begin{eqnarray}}
\def \eeqna{\end{eqnarray}}
\def \beq{\begin{equation}}
\def \eeq{\end{equation}}

\def \ol{\overline}

\def \b{\beta}

\def \sig{{\bf\sigma}}

\def \ab2{\alpha\beta^2}

 \newcommand \be {\begin{equation}}
\newcommand \bea {\begin{eqnarray} \nonumber }
\newcommand \ee {\end{equation}}
\newcommand \eea {\end{eqnarray}}
 \newcommand \eps {\epsilon}

\newcommand \G {\Gamma}

\input{epsf}
\begin{document}
\twocolumn[\hsize\textwidth\columnwidth\hsize\csname@twocolumnfalse\endcsname
\preprint{MA/UC3M/11/95}
\title{Phase diagram of glassy systems in an external field}

\author{ Silvio Franz(*) and Giorgio Parisi(**)}
\address{
(*) International Center for Theoretical Physics\\
Strada Costiera 11,
P.O. Box 563,
34100 Trieste (Italy)\\
(**) Universit\`a di Roma ``La Sapienza''\\
Piazzale A. Moro 2, 00185 Rome (Italy)\\
e-mail: {\it franz@ictp.trieste.it, parisi@roma1.infn.it}}
\date{December 1996}
\maketitle

\begin{abstract}
We study the mean-field phase diagram of glassy systems in a field 
pointing in the direction of a metastable state. We find competition
among a ``magnetized'' and a ``disordered'' phase, that are separated by 
a coexistence line as in ordinary first order phase transitions. 
The coexistence line terminates in a critical point, which in principle 
can be observed in numerical simulations of glassy models. 
\end{abstract}
\twocolumn
\vskip.5pc] 
\narrowtext 
The crossover from supercooled liquids to glasses, although manifestly 
an off-equilibrium  phenomenon, presents 
characteristics reminiscent of equilibrium phase transitions \cite{vetro}.
Probed at fixed observation time scale, the various 
thermodynamic quantities show non-analytic behavior at the
glassy temperature $T_g$. While the singularity is pushed to lower 
temperatures for increasing observation times, it does not seen to
decrease in intensity. Whether for infinite probing times
the singularity persists to produce a ``true'' finite temperature
phase transition is a longly debated question, and is at present
experimentally and theoretically unsolved. Experimentally, due to the 
rapid growth of the relaxation time as temperature is lowered, 
it is difficult to equilibrate the systems at temperatures
that exceed  the estimated critical temperature by less then 15 \%.
Theoretically, the only well established issues rely on mean-field 
approximation.
Mean field theory for disordered models that are conjectured to be in 
the same universality class of structural glasses, predicts a thermodynamical
transition, along the Gibbs-DiMarzio scenario of the entropy crisis at 
finite temperature $T_s$ \cite{kirtir}. 
The transition at $T_s$ combines  features of first order and second order 
phase transitions.
Like in a second order phase transition the energy is continuous and
the specific heat has a jump. Like in a first order phase transition
there is no divergent susceptibility. 
Detailed studies of the mentioned models \cite{kirtir}
have shown that it exists an interval of temperatures $T_s<T<T_d$ where 
the thermodynamic, as well as the dynamic is dominated by the existence 
metastable states, which can trap the system for infinite time. 
In finite dimension  the picture must necessarily be modified, as 
metastable states can not have a infinite time life. An appealing possibility 
is that  metastable states continue to exist with finite decay time. 
Transitions among metastable states would dominate the dynamics 
in that region. The inclusion of these phenomena 
is one of the most challenging problem in the theory of glasses \cite{parisi}. 

In ordinary mean-field theories the appearance of 
metastable states is signaled by  local minima 
of the ``potential function'',
the free-energy as a function of the order parameter,
and can be stabilized 
by the introduction of an external field coupled with the order parameter. 
In glasses, while the system is dynamically frozen,  
long range order is absent and it is impossible to identify 
any intrinsic (physical) order parameter for a single system allowing
to distinguish different thermodynamical states.
The usual Edwards-Anderson parameter, that measure the degree 
of freezing, takes the same value in all the dominant states. 
Spin glass theory \cite{MPV}
suggests that measures of distance (or alternatively 
similarity) among different ``real replicas'' can be taken as appropriate 
order parameter. A potential, or better a class of potential functions, 
can be defined considering the partition function of several real 
replicas with constraints on the mutual distances \cite{FPVI}. 
Alternatively one can release the constraints and introduce a field coupled
to the distances. The two constructions
are related by the Legendre transform, and both
have been employed to study glassy systems \cite{FPVI,KPV,remi,I}.
This point of view, although it can not help in real experiments,
can give theoretical insights, and can be useful in numerical experiments.
Different question can be addressed with different potential function. 
In this paper we limit ourselves to potentials involving two real replicas. 
More complicated three-replica potential have been considered in 
\cite{FPVII,AIG}.
Let us, to be defined, consider a system of $N$ interacting variables $S_i$,
$i=1,...,N$, with Hamiltonian $H(\vS)$, and be $q(\vS,\vS')$ an overlap
function, i.e. an intensive 
measure of similarity among the configurations 
$\vS$ and $\vS'$. 
We can consider two replicas of the system trying to thermalize 
at the same temperature with an attractive coupling. 
The appropriate thermodynamical description of this situation is given
by the partition sum:
\be
Z_A =
\sum_{\vS_1,\vS_2} \exp\left[-\b( H(\vS_1)+H(\vS_2)) 
+\b\eps N q(\vS,\vS')\right].
\ee
We will call $\G_A(\eps)=-T/N \log Z_A$ the annealed potential. 
Alternatively, we can consider a reference configuration $\sig$ thermalized at 
temperature $T'$, acting as an external field on a replica $\vS$ which
tries to thermalize at temperature $T$. In this case the right partition
function is 
\be 
Z_Q=
\sum_{\vS} \exp\left[-\b H(\vS)
+\b\eps N q(\vS,\sig)\right],
\label{zq}
\ee
and 
$\G_Q(\eps)=-T/N \log Z_Q$ will be the quenched  potential. 
The idea of the quenched potential is particularly appealing 
for the physics of glasses. 
At a given cooling rate, the system equilibrates within the super-cooled
liquid phase, until the glassy transition
temperature $T_g$ is reached. Below that 
temperature 
it is reasonable to think that the system will be stacked for a long 
time in a metastable state. While fast degree of freedom 
thermalize at temperature $T$, the slow degree of freedom will remain 
confined for a long time close to the equilibrium  configuration 
reached at $T_g$. In such conditions,  restricted Boltzmann-Gibbs
averages generated by a partition function of the kind (\ref{zq})
with $T'$ identified with $T_g$ could be appropriate. In this
pseudo-thermodynamic description the parameter $\eps$, which fixes the
equilibrium value of the overlap with the reference configuration, 
should be considered as a dynamical variable, 
fixed by the condition that the system has equilibrated in the 
neighbors of the reference configuration $\sig$, but has not 
had the time necessary to escape from the metastable state.  
Within the limit of validity of this picture the usual laws 
of thermodynamics must hold in the glassy phase. The picture certainly loose
its validity for times large enough to escape from the metastable state
and aging dynamics sets in the system \cite{struik}.

The quenched case in an 
infinitesimal field and $T=T'$ was studied in detail in 
\cite{remi}. In this paper we extend the analysis
to  the equilibrium
phase diagram of the quenched potential in the $T-\eps$ plane. 
For definiteness let us introduce the model 
that we will use in the following, the same lines of reasoning can
be followed in general.
The model we study is a spherical $p$-spin model \cite{pspin}. 
This is defined in terms of
$N$ real dynamical variables $S_i$, ($i=1,...,N$) subjected to the 
constraint $\sum_{i=1}^N S_i^2=N$ and 
interacting via a random 
Gaussian Hamiltonian $H(\vS)$, defined by its correlator
$\ol{H(\vS)\; H(\vS')}=N f(q_{\vS,\vS'})$. The overlap
$q_{\vS,\vS'}$ is a measure of similarity among 
configurations and is given by  $q_{\vS,\vS'}=1/N \sum_i S_i S_i'$. 
If the correlation function is  taken of the monomial form $f(q)=1/2 q^p$
the Hamiltonian can be represented as $H=-\sum_{i_1<...<i_p}^{1,N}
J_{i_1,...,i_p}S_{i_1}...S_{i_p}$ with independent centered Gaussian 
couplings $J_{i_1,...,i_p}$, which motivates the name of the model.
We will stick to this case throughout all this paper.  
The model has been studied extensively during the last years, and 
furnishes, for $p>2$, a good mean-field model of fragile glass. 
It has been often observed \cite{kirtir} 
that the Langevin relaxation of this model 
leads to equations homologous to those of mode coupling theory \cite{MCT}.

The quenched free energy $\Gamma_Q[{\bf \sig};\b,\eps]$ should
be self-averaging with respect to the distribution of $\sig$, so 
we can compute it as 
$\Gamma(\eps,T,T')=-1/Z \sum_{\sig}\exp(-\b' H(\sig))T/N 
\log Z[{\bf \sig};\eps]$.
In the following we will study the phase diagram of the model in
the $\eps -T$ plane in two situations: a) $T'=T$, corresponding to 
restricting the partition sum to the vicinity of a particular equilibrium 
state at each temperature. b) $T'$ fixed, corresponding to probe 
the evolution of the free-energy landscape in the vicinity of 
a fixed configuration of equilibrium at $T'$ when $T$ is changed. 
 
The Legendre transform of $\G(\eps,T,T')$, 
$V(q,T,T')\equiv\min_\eps \G(\eps)+\eps q$, which corresponds physically to
constraining the value of $q_{\vS,\sig}$ to $q$, was studied in detail in 
\cite{I} with the aid of the replica method. The interested reader 
can found there
 details about the general method and the analytic expression for 
$V$ in the case of the spherical $p$-spin model. 

\begin{figure}
\epsfxsize=250pt
\epsffile{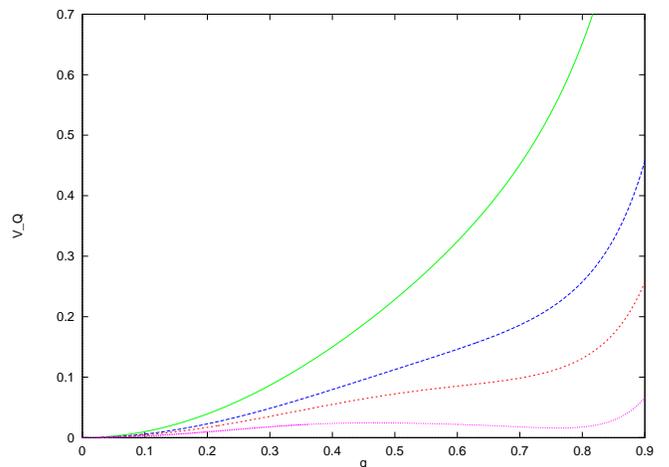}
  \caption[0]{\protect\label{F_E2}
 The potential as a function of $q$ for $T'=2 T_s T_d/(T_s+T_d)$ and various
values of $T$, in order of decreasing temperatures from top to bottom. 
  }
\end{figure}
The shape of the function $V$ turned out to be the characteristic one
of a mean-field system undergoing a first order phase transition. 
At high enough temperature $V$ is an increasing and convex function of 
$q$ with a single minimum for $q=0$. Decreasing the temperature to a value 
$T_f$, where for the first time a point $q_f$ with $V''(q_f)=0$ appears,
the potential looses the convexity property and a phase transition 
can be induced by a field. A secondary minimum develops at $T_d$,
the temperature of dynamical transition \cite{kirtir},  signaling
the presence of long-life metastable states.
The minimum of the potential has received a dynamical interpretation
\cite{I,BBM} as corresponding to the states reached at long times 
by the evolution at temperature
$T$ starting at time zero from an equilibrium configuration at temperature
$T'$. This allows to fix $q$ in the pseudo-thermodynamical 
description mentioned above
to the value it takes in the minimum of the potential. 
The height 
of the secondary minimum reaches the one of the primary minimum
at $T=T_s$  and coexistence in zero field
takes place. This is the usual  statical transition point in zero field,
and it is not accompanied by the release of latent heat. 
In figure 1 we show the shape of the potential in the various regions. 
The attentive reader would have noticed that respect to the curves presented 
in \cite{I} only one secondary minimum is present at low temperature. The 
results 
we present here are corrected taking into account replica 
symmetry breaking effects, the meaning of which will be discussed in \cite{me}.
Although the behavior of the potential 
function is analogous to the one found in ordinary systems 
undergoing a first order phase transition
the interpretation is here radically different. While in ordinary cases 
different minima represent 
 qualitatively different  thermodynamical states (e.g. gas and liquid), 
this is not the case in the potential discussed here. 
In our problem the local minimum appears when ergodicity is broken, and the 
configuration space  splits into an exponentially large number 
of components. The two minima are different manifestations of 
states with the same characteristics. The height of the secondary minimum, 
relative to the one at $q=0$ measures the free-energy loss to keep the 
system in the same component of the quenched one. At equal temperatures 
$T=T'$ this is just the complexity $T\Sigma$, i.e. the logarithm of 
the number of distinct pure states of the system.
For $T\neq T'$ it  also takes into account the free-energy variation
of the equilibrium state at temperature $T'$ when followed at temperature $T$. 
\begin{figure}
\epsfxsize=250pt
\epsffile{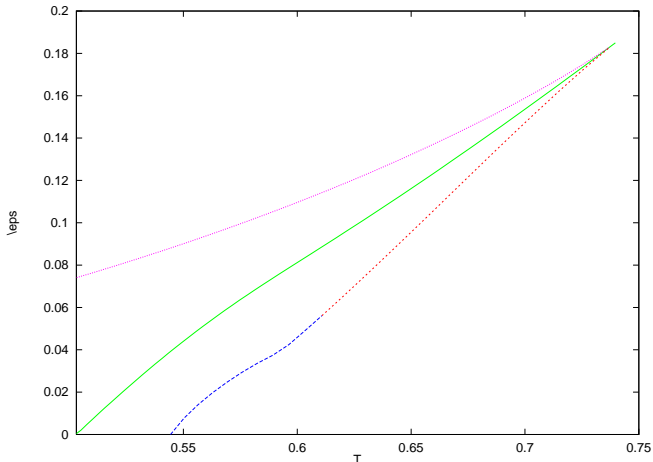}
  \caption[0]{\protect\label{F_E1}
Phase diagram in the $\eps-T$ plane for $T'=T$. 
The upper curve is the spinodal of the disordered state, the 
lower one the spinodal of the magnetized  state, and the middle curve
the coexistence line. The coexistence line touches the axes $\eps=0$ at
$T=T_s$, while the spinodal of the magnetized state touches it at
$T=T-d$. For $T<T_s$ the spinodal of the disordered state 
remains finite and touches the $T=0$ axes at finite $\eps$. 
  }
\end{figure}

The presence of the field $\eps$ adds finite stability to the 
metastable state,
and the transition is displaced at higher 
temperatures.
 In figure 2 we display the phase diagram of the $p=4$ model 
in the case $T'=T$. The coexistence line departs from the axes $\eps=0$ 
at the transition temperature $T_s$ and reaches monotonically a critical point 
$(T_{cr},\eps_{cr})$. In figure 2 it is also shown the spinodal of the 
``magnetized'' solution, which touches the $\eps=0$ axes at the dynamical 
temperature $T_d$, and the spinodal of the ``disordered'' solution 
for temperaratures larger then $T_s$. 

 The coexistence line for $T'$ fixed 
in the interval $T_d\leq T'\leq T_s$
is qualitatively similar to the one of figure 2 at high 
enough temperature, but (for $T'> T_s$) it never touches the axes $\eps=0$.
\begin{figure}
\epsfxsize=250pt
\epsffile{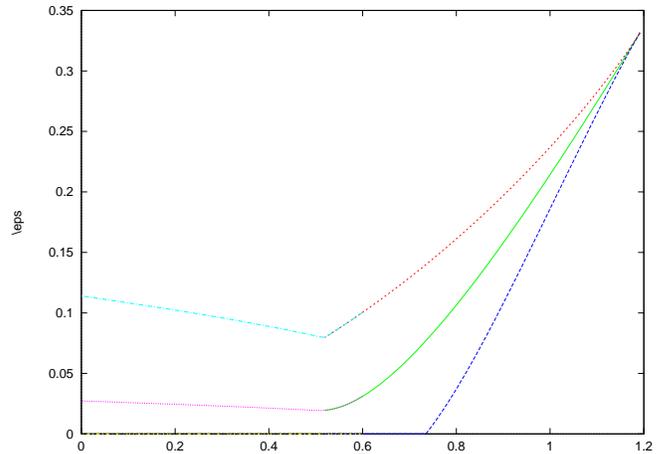}
  \caption[0]{\protect\label{F_E3}
Phase diagram in the $\eps-T$ plane for $T'=2 T_s T_d/(T_s+T_d)$. 
The upper curve is the spinodal of the disordered state, the 
lower one the spinodal of the magnetized  state, and the middle curve
the coexistence line. 
  }
\end{figure} 
Even at zero temperature there is a first order phase transition in $\eps$,
reflecting the fact that the ground state of the system is lower then the 
energy of the
reference state when followed at $T=0$. This can be seen in figure 3 where we 
show the phase diagram for $T'=2 T_s T_d/(T_s+T_d)$. 
At the critical point the transition is second order.
\begin{figure}
\epsfxsize=250pt
\epsffile{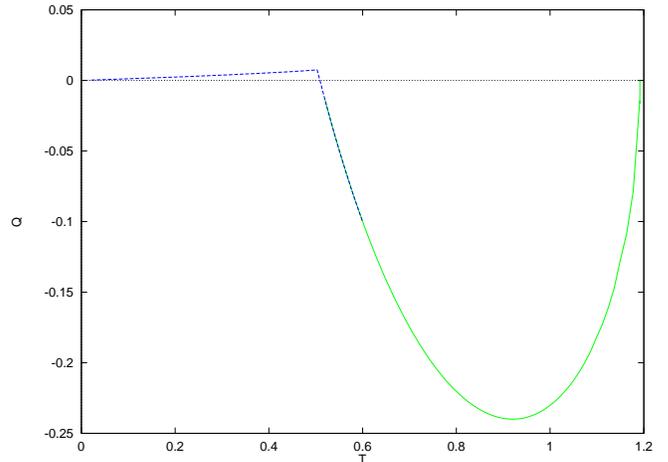}
  \caption[0]{\protect\label{F_E4}
Latent heat of the transition as a function of the temperature 
for $T'$ as in fig. 3.
  }
\end{figure}
While the transition in zero field is not accompanied by 
 heat release, a latent heat is present in non zero $\eps$.
In
figure 4 we show, in the same conditions of fig. 3,
 the latent heat ${\cal Q}=E_+-E_- -\eps(q_+-q_-)$ 
where $E_+$ ($q_+$) and 
$E_-$ ($q_-$)
are the internal energies (overlaps) respectively of the magnetized and 
unmagnetized states.
Notice that (as it should) the latent heat is zero at
the critical point and at $T=0$. The magnetized state
roughly reflects  the properties of the equilibrium states
at temperature $T'$ followed at temperature $T$, while 
the disordered state  reflects  the properties of  the true
equilibrium states at temperature $T$. 
We see that at high temperature 
the magnetized state is energetically favored, while at low temperature
it has an energy higher then the one of equilibrium. 
The  point 
where ${\cal Q}$ changes sign does not correspond to a second order 
phase transition. Finally, in figure 5 we show, for a fixed temperature
the curve of $q(\eps)$ obtained by the Maxwell construction. 
\begin{figure}
\epsfxsize=250pt
\epsffile{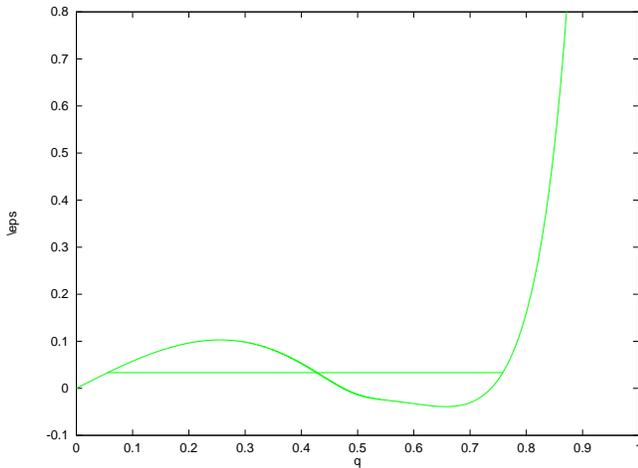}
  \caption[0]{\protect\label{F_E5}
Equation of state for $T'=2 T_s T_d/(T_s+T_d)$ and $T=0.609$. 
The horizontal line
corresponds to coexistence and is obtained by the Maxwell construction.
  }
\end{figure}
So far we have presented results for the 
quenched potential $\G_Q$. 
In the annealed case, it was shown in \cite{KPV} the potential 
$V_a$ has qualitatively the same shape than the quenched one ($V_Q$); 
it can therefore be expected  a phase diagram similar to
the one of figure 2 also in this case. 

Although we have based our discussion on a mean-field model, 
we expect that
the qualitative 
features of the phase diagrams presented survive in finite dimension. 
We belove that the existence of a coexistence line, terminating in
a critical point, is a constitutive feature of systems whose physics
is dominated by the existence of long lived metastable states. 
The predictions of 
this paper can be submitted to numerical test in glassy model 
systems as like e.g. Lennard-Jones or hard spheres, 
or polymer glasses. For example the identification of the
complexity $\Sigma$ as the free energy difference 
between the stable and the metastable phases
allows a direct measure of this quantity in a simulation. Indeed 
the ending of the transition lines in a critical point 
implies that the metastable state can be reached via closed paths 
in phase diagram  leaving always the system  in (stable or metastable)
equilibrium; and the free energy difference of the two phases 
computed integrating the specific heat a long the loop.

{\bf Acknowledgments}

The authors benefited of the stimulating environments of the
conferences of Sitges (10 - 14 June 1996) and Lyon 
(30 September - 3 October 1996).
SF thanks the ``Dipartimento di Fisica dell' Universit\`a di Roma
La Sapienza'' for kind hospitality during the elaboration of this work.

\end{document}